\documentclass[a4paper,11pt]{article}
\usepackage{physics}
\usepackage{graphicx}
\usepackage{booktabs}
\usepackage{bbold}
\usepackage{bm}
\usepackage{pos}
\usepackage{tikz}

\usepackage[compat=1.1.0]{tikz-feynman}
\usepackage[outline]{contour}
\contourlength{3pt}
\usetikzlibrary{decorations.markings}
\usetikzlibrary{arrows.meta,positioning}
\tikzset{%
->-/.style={decoration={markings,mark=at position #1 with {\arrow[fill=black]{>}}},
            postaction={decorate}
            },
->-/.default=0.5
}

\title{Towards the time-like pion form factor beyond the elastic regime using domain-wall QCD}
\ShortTitle{Towards the time-like PFF beyond the elastic regime using DW-QCD}

\author*[a,b]{Gabriele Morandi}
\author[a,b]{Mattia Bruno}
\author[a,b]{Francesca Argia Bresciani}
\author[c]{Christoph Lehner}
\author[c]{Julian Parrino}

\affiliation[a]{Dipartimento di Fisica, Universit\`a degli Studi di Milano-Bicocca, \\ Piazza della Scienza 3, I-20126 Milano, Italy}
\affiliation[b]{INFN, Sezione di Milano-Bicocca, \\ Piazza della Scienza 3, I-20126 Milano, Italy}
\affiliation[c]{Fakult\"at f\"ur Physik, Universit\"at Regensburg, \\ Universit\"atsstraße 31, 93040 Regensburg, Germany}

\emailAdd{gabriele.morandi@unimib.it}
\emailAdd{mattia.bruno@unimib.it}
\emailAdd{f.bresciani4@campus.unimib.it}
\emailAdd{christoph.lehner@ur.de}
\emailAdd{julian.parrino@ur.de}

\abstract{In this work, we investigate the time-like pion form factor from lattice QCD in the isosymmetric limit, a quantity that plays an important role in understanding hadron physics with substantial phenomenological applications. This observable can be calculated in the elastic region using the finite-volume approach, up to the first (four-particle) open channel. With the goal of accessing the exclusive two-pion form factor in the inelastic region, starting from a three-point correlator involving the vector current and two (temporally-displaced) pion interpolating operators, we examine the associated underlying spectral density and calculate the form factor using a formalism based on the LSZ reduction.
A preliminary analysis on one ensemble generated by the RBC/UKQCD collaboration using domain-wall fermions is presented.}

\FullConference{The 42nd International Symposium on Lattice Field Theory (LATTICE2025)\\
2-8 November 2025\\
Tata Institute of Fundamental Research, Mumbai, India\\}

\begin{document}
\maketitle

\section{Introduction}

The pion form factor $F_\pi$ (PFF) is a physical quantity which encodes information about the internal structure of the pion and its interaction with the electromagnetic field. 
It plays a central role in hadron phenomenology: in the time-like region it is dominated by hadronic resonances—most notably the $\rho$ meson—while more generally it enters precision observables such as the hadronic vacuum polarization contribution to the muon $g-2$ \cite{Aoyama:2020ynm,Aliberti:2025beg}, as well as the pion charge radius \cite{Feng:2019geu}. As it is controlled by the QCD dynamics in the strong-coupling regime, its determination from first principles must be carried out using non-perturbative methods, such as Lattice QCD simulations.

State-of-the-art calculations rely on the L\"uscher finite-volume method \cite{Luscher:1986pf,Luscher:1990ux}, in which lattice-accessible finite-volume matrix elements are related to the corresponding infinite-volume scattering amplitudes by means of a determinant equation. In particular, starting from the lattice determinations of the energy spectrum and matrix elements of the vector current, the infinite-volume form factor is determined by multiplying the latter by the appropriate Lellouch-L\"uscher factor \cite{Lellouch:2000pv, Meyer:2011um}; notable examples of these calculations can be found in Refs.~\cite{Feng:2014gba,Bulava:2015qjz,Erben:2019nmx}. 
However, this methodology is limited to the elastic $\pi \pi$ kinematical region, which shrinks as the pion mass is lowered. In our study, the first relevant open channel is the four-particle one, and therefore we do not profit from the recent advances in the three-particle quantization condition~\cite{Polejaeva:2012ut,Briceno:2012rv,Hansen:2014eka}. 
In addition, further complications arise within this formalism. 
The relevant determinant equations are, in principle, infinite dimensional relations which must be truncated to the first lowest partial waves of the total-momentum little group, introducing systematic uncertainties (and limiting sensitivity to higher angular momenta). 
Moreover, as the spatial volume increases, the gap among energy levels in the spectrum and the differences with the corresponding non-interacting states---essential to the analysis—become smaller in magnitude, making the extraction of the corresponding infinite-volume quantities increasingly challenging.
Furthermore, each quantization condition generally couples multiple scattering channels, so isolating specific physical processes requires global model parametrizations. Consequently, at the physical point, resonances already lying in the inelastic regime, such as the $\rho$ meson, require a careful treatment of the associated systematic errors with this methodology. 

Given these limitations, it is worthwhile exploring alternative approaches to access the time-like PFF beyond the inelastic threshold. As first outlined in Ref.~\cite{Hansen:2017mnd} and then further developed in Refs.~\cite{Bulava:2019kbi,Bruno:2020kyl,Patella:2024cto}, lattice Euclidean correlation functions can be related to the corresponding finite-volume spectral densities, which in turn are the objects relevant for extracting infinite-volume amplitudes. In particular, the temporal separation between the operators in the correlator plays a crucial role in isolating the relevant states, while the smearing of the spectral density with a suitably chosen kernel is a key ingredient for obtaining the desired amplitudes. 

In this work, we present an exploratory study of a recently proposed methodology, with the ultimate goal of extracting $F_\pi(q^2)$ across the full kinematically allowed energy range. 
This approach primarily serves as a test bed for the application of  inverse methods in three-point functions in QCD, paving the way for more appealing applications to other phenomenologically relevant processes. A notable example is the study of hadronic weak decays into two-hadron final states (relevant in flavor-physics studies): in this situation one would need a four-point function adding an extra layer of complication, together with the possible non-trivial renormalization of the corresponding weak operator. In our study, we simplify these aspects by considering a current protected by a Ward Identity and one external vacuum state. 

This work is organized as follows: in Section~\ref{sec:methodology} we introduce the main objects of this study and provide a brief recap of how Euclidean lattice correlators can be related to infinite-volume scattering amplitudes; Section~\ref{sec:results} presents preliminary results from our ongoing calculation performed on an ensemble generated by the RBC/UKQCD collaborations. Finally, in Section~\ref{sec:conclusions} we summarize our findings and discuss future perspectives.

\section{Methodology}\label{sec:methodology}

We focus on the determination of the time-like PFF in the center-of-mass frame using the isospin 1 local electromagnetic vector current, namely
\begin{equation}
    V_k(x) = \frac{1}{2} \left\{ \bar{u} \gamma_k u - \bar{d} \gamma_k d \right\}(x) \,.
\end{equation}
The chosen current is tailored to project onto the two-pion $p$-wave state.  
In this kinematical configuration, their total spatial momentum vanishes and their momenta are exactly back-to-back. 
By taking the Dirac matrices in Euclidean space, we define the PFF from the following matrix element
\begin{equation}
    \mel{0}{V_k(q)}{\pi \pi, \vb{0}} = 2 \vb{p}_k F_\pi(4 \omega_{\vb{p}}^2) \,, \quad q \equiv p_1+p_2 \,, \quad k = 1,2,3 \,,
\end{equation}
where $p_1$ and $p_2$ are the four momenta of the pions and $\omega_{\vb{p}} = \sqrt{m_\pi^2 + \vb{p}^2}$ is the pion energy. \\

\begin{figure}[h]
    \centering
    \begin{minipage}{0.485\textwidth}
        \centering
        \begin{tikzpicture}[scale=0.55]
            \draw[thick,->-] (5.5,3) to [bend left] (1.5,3);
            \draw[thick,->-] (1.5,3) to [bend left] (3.5,4.15);
            \draw[thick,->-] (3.5,4.15) to [bend left] (5.5,3);

            \draw[fill=black] (3.5,4.15) circle [radius=2pt];
            \draw[fill=black] (1.5,3) circle [radius=2pt];
            \draw[fill=black] (5.5,3) circle [radius=2pt];

            \draw (1.5,2.8) node[left] {\contour{white}{$V_k(\tau;\vb{0})$}};
            \draw (5.5,2.8) node[right] {\contour{white}{$\mathcal{O}_{\pi,2}^\dag(0)$}};
            \draw (3.5,4.2) node[above] {\contour{white}{$\mathcal{O}_{\pi,1}(t;\vb{p})$}};
        \end{tikzpicture}
    \end{minipage}
    \hfill
    \begin{minipage}{0.485\textwidth}
        \centering
        \begin{tikzpicture}[scale=0.23]
            \draw[thick,->-] (9.5,3) to [bend left] (-1.5,3);
            \draw[thick,->-] (-1.5,3) to [bend left] (0.5,5.8);
            \draw[thick,->-] (0.5,5.8) to [bend left] (9.5,3);
    
            \draw[fill=black] (0.5,5.8) circle [radius=4pt];
            \draw[fill=black] (-1.5,3) circle [radius=4pt];
            \draw[fill=black] (9.5,3) circle [radius=4pt];
    
            \draw (-1.5,2.8) node[left] {\contour{white}{$V_k(\tau;\vb{0})$}};
            \draw (9.7,2.8) node[right] {\contour{white}{$\mathcal{O}_{\pi,2}^\dag(0) \, \frac{e^{E_\pi(\vb{p})t}}{Z_{\pi,2}^{1/2}(\vb{p})}$}};
            \draw (-2.8,5.5) node[above] {\contour{white}{$\mathcal{O}_{\pi,1}(t;\vb{p})$}};
        \end{tikzpicture}
    \end{minipage}
    \caption{\textit{Left}: three-point function diagram. The vector current is inserted at time-slice $\tau$ and the pion operators are temporally displaced. \textit{Right}: sketch of the extraction of the correlator $G_k(\tau-t; \vb{p})$ by looking at the limits $t,\tau \to \infty$, once the operator dependence on $\mathcal{O}^\dag_{\pi,2}(0)$ is removed.}
    \label{fig:3pt_diagrams}
\end{figure}
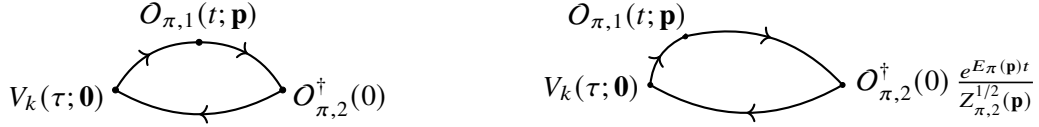

The relevant Euclidean three-point correlation function in our study is \begin{equation}\label{eq:def_3pt}
    C_k(t,\tau; \vb{p}) \equiv \big\langle V_k(\tau; 0) \mathcal{O}_{\pi,1}(t; \vb{p}) \mathcal{O}_{\pi,2}^\dag(0) \big\rangle \,, \quad k = 1,2,3 \,,
\end{equation}
where the interpolating operators in time-momentum representation are defined as
\begin{align}
    V_k(x_4; \vb{p}) &= \sum_{\vb{x}} e^{-i \vb{p} \cdot \vb{x}} V_k(x) \,, \\
    \mathcal{O}_{\pi,i}(x_4; \vb{p}) &= \sum_{\vb{x}} e^{-i \vb{p} \cdot \vb{x}} \mathcal{O}_{\pi,i}(x) \,, \quad i = 1,2 \,.
\end{align}
The operators $\mathcal{O}_{\pi,i}(x)$ are both of the form $\left( \bar{u} \Gamma_5 d \right)(x)$, with $\Gamma_5$ denoting an appropriate Dirac matrix carrying the quantum numbers of the pion. We remark that the two interpolating pion operators are temporally displaced. The corresponding non-vanishing Wick contractions are diagrammatically represented in the left panel of Fig.~\ref{fig:3pt_diagrams}.

Now, let us consider the spectral decomposition of the correlator in Eq.~\eqref{eq:def_3pt}. Following the prescription outlined in Refs.~\cite{Bulava:2019kbi,Bruno:2020kyl}, we insert a complete set of states $\mathbb{I} = \sum_{n,\vb{k}} \ket{n,\vb{k}} \! \bra{n,\vb{k}}$ between the pion operators to isolate the temporal dependence of the remaining two-point correlator involving $V_k$ and $\mathcal{O}_{\pi,1}$. This yields\footnote{The notation $\mathcal{O}(t)$, for generic operators $\mathcal{O}$, indicates that only the momentum dependence in $\mathcal{O}(t; \vb{p})$ has been extracted using the Heisenberg representation.} 
\begin{equation}
    C_k(t,\tau; \vb{p}) = \sum_{n} e^{-E_n(\vb{p}) t} \mel{0}{V_k(\tau) \mathcal{O}_{\pi,1}(t)}{n,\vb{p}} \! \mel{n,\vb{p}}{ \mathcal{O}_{\pi,2}^\dag(0)}{0} \,, 
\end{equation}
where $E_n(\vb{p})$ is the finite-volume energy of the state $\ket{n, \vb{p}}$. Notice that the Heisenberg representation of the operators has been exploited to select in the sum only the states with total spatial momentum $\vb{p}$. Let us focus on the limits for $\tau, t \to \infty$ in order to isolate the dominant contribution given by the single-pion state, namely $\ket{n, \vb{p}} = \ket{\pi, \vb{p}}$. Neglecting thermal corrections and effects from heavier energy states\footnote{Such effects would appear via a dependence on $\tau$.}, the spectral decomposition can be expressed in the form
\begin{equation}\label{eq:3pt_lim_t}
    C_k(t,\tau; \vb{p}) \simeq Z_{\pi,2}^{1/2}(\vb{p}) e^{-E_\pi(\vb{p}) t} G_k(\tau-t; \vb{p}) \,, \quad t, \tau \gg 0 \,,
\end{equation}
where $Z_{\pi,i}^{1/2}(\vb{p}) \equiv \mel{\pi, \vb{p}}{\mathcal{O}_{\pi,i}^\dag(0)}{0}$. The two point correlator $G_k$ is defined as
\begin{equation}\label{eq:G2_def}
    G_k(\tau-t; \vb{p}) \equiv \mel{0}{V_k(0) e^{-\widehat{H} (\tau-t)} \mathcal{O}_{\pi,1}(0)}{\pi, \vb{p}} \,,
\end{equation}
with $\widehat{H}$ being the QCD Hamiltonian, and it admits the following spectral representation
\begin{equation}\label{eq:spec_dec_G2}
    G_k(\tau-t; \vb{p}) = \sum_{n} e^{-E_n(\vb{0}) (\tau-t)} \mel{0}{V_k(0)}{n,\vb{0}} \! \mel{n,\vb{0}}{\mathcal{O}_{\pi,1}(0)}{\pi, \vb{p}} \,.
\end{equation}
Notice that, given our choice of the momenta in $C_k(t,\tau; \vb{p})$, the intermediate states are all at rest. 
The calculation of the correlator $G_k$ is the main focus of this work, as it encodes the relevant information regarding the time-like PFF. In fact, as pointed out in Ref.~\cite{Bulava:2019kbi}, once the finite-volume smeared spectral density 
\begin{equation}\label{eq:rho_def}
    \rho_{k,\vb{p}}(\mathcal{E} | \sigma, \mathcal{O}_{\pi, 1}) \equiv \sum_{n} \frac{i}{\mathcal{E} - E_n(\vb{0}) + i \sigma} \mel{0}{V_k(0)}{n,\vb{0}} \! \mel{n,\vb{0}}{\mathcal{O}_{\pi,1}(0)}{\pi, \vb{p}} \,,
\end{equation}
is computed from $G_k$, the corresponding on-shell scattering amplitude at $\mathcal{E} = 2 E_\pi(\vb{p})$ is obtained by applying the LSZ reduction formula and by studying the residue at the pole of the spectral density. Specifically, the time-like pion form factor of interest is extracted through the following ordered double limit \cite{Bulava:2019kbi}
\begin{equation}\label{eq:master_formula}
    \lim_{\sigma \to 0^+} \lim_{L \to \infty} \frac{2 E_\pi(\vb{p})}{Z_{\pi,1}^{1/2}(\vb{p})} \sigma \rho_{k,\vb{p}}(\mathcal{E} | \sigma, \mathcal{O}_{\pi, 1}) = 2 \vb{p}_k F_\pi(\mathcal{E}^2) \,, \quad \text{for} \quad \mathcal{E} = 2 E_\pi(\vb{p}) \,.
\end{equation}
Given the results in Eqs.~\eqref{eq:rho_def} and \eqref{eq:master_formula}, several remarks are in order. First, the sum entering the spectral density $\rho_{k,\vb{p}}$ could, in principle, be implemented directly once the finite-volume energy spectrum and the matrix elements are known. However, a naive multi-exponential fit of $G_k$ is affected by the signal-to-noise problem at large $t$'s, making the extraction of many states unfeasible. Instead, the adopted strategy was to exploit a dedicated GEVP \cite{Luscher:1990ck,Blossier:2009kd} analysis to first determine the spectrum (and the matrix elements involving $V_k$), and then to perform a constrained fit of $G_k$ in order to extract the remaining matrix elements of $\mathcal{O}_{\pi,1}$. Alternatively, the spectral density could be reconstructed directly from $G_k$ using suitable inverse-problem techniques, such as Backus-Gilbert \cite{Backus:1968}-type methods. The former strategy is further described in Section~\ref{sec:results}, while the latter is currently under investigation. 

A second aspect concerns the ordering of the limits, which is essential for correctly recovering the infinite-volume amplitude. In a finite box, the spectral density is given by a sum of delta functions, and becomes a smooth function only in the limit $L \to \infty$. Hence, the smearing parameter $\sigma$ should remain sufficiently large to avoid singularities in the spectral density and to guarantee small finite-size effects; at the same time, $\sigma$ should be sufficiently small—and ultimately taken to zero—to ensure the validity of the LSZ prescription described in Ref.~\cite{Bulava:2019kbi}. The two conditions together define a typical window problem. For a zero to two transition, such as the process considered here, only a single pole must be amputated in $\rho_{k,\vb{p}}$ at the on-shell energy, which accounts for the additional factor of $\sigma$ appearing in front. Notably, once the operator dependence given by $\mathcal{O}_{\pi,1}$ is removed, the LSZ residue on the l.h.s.~is expected to be unique. 
This will be an important self-consistency cross-check of the calculation, c.f. Fig.~\ref{fig:reconstruction_rho}. 

Finally, in contrast to standard finite-volume approaches, no assumptions about the dynamics of open scattering channels are required, since, in principle, any energy region can be accessed at the on-shell point $2 E_\pi(\vb{p})$, allowing to probe the PFF beyond the inelastic threshold. In practice, an energy scan of the PFF is achieved by varying the pion momentum $\vb{p}$ and in our preliminary study we test the validity of the method at the first accessible value of $\vb{p}$. On top of that, both the real and imaginary parts of $F_\pi$ can be independently determined by selecting the corresponding real or imaginary components of the smearing kernel in Eq.~\eqref{eq:rho_def}.

\section{Results}\label{sec:results}

In this study the relevant Wick contractions have been evaluated on $50$ previously generated RBC/UKQCD \cite{RBC:2014ntl} gauge field configurations, at physical pion mass with $2+1$ flavors of M\"obius domain-wall fermions. The ensemble parameters are summarized in Tab.~\ref{tab:ensemble}.

\begin{table}[ht]
    \centering
    \begin{tabular}{ccccccc}
        \toprule
        Size & $\beta$ & $m_\pi[\text{MeV}]$ & $m_K[\text{MeV}]$ & $a^{-1}[\text{GeV}]$ & $L[\text{fm}]$ & $m_\pi L$ \\
        \midrule
        $48^3 \times 96 \times 24$ & $2.13$ & $139.2(4)$ & $497.33(54)$ & $1.730(4)$ & $5.476(12)$ & $3.863(6)$ \\[.5ex]
        \bottomrule
    \end{tabular}
    \caption{Input parameters and relevant quantities of the ensemble \texttt{48I} \cite{RBC:2014ntl}, where the size of the lattice is written in the form $L^3 \times T \times L_s$, with $L$ and $T$ being the spatial and temporal lattice extent in lattice units, respectively; $L_s$ corresponds to the extent of the fifth dimension in lattice units.}
    \label{tab:ensemble}
\end{table}

The numerical results presented in this section are obtained from a preliminary analysis of the correlator $C_k(t,\tau; \vb{p})$ for $\tau/a = \{12, 24, 36\}$ and momenta such that $(\frac{L}{2 \pi})^2 \abs{\vec{p}}^2 = 1$. 
For the correlator we start with a point source in $\mathcal{O}^\dag_{\pi,2}(0)$, following a sequential inversion at $V_k(\tau; \vb{0})$, before closing the trace using $\gamma_5$-hermiticity.
The propagator inversions in the calculation of $C_k$ have been performed using equally distributed point sources on a grid of $64$ points (see for instance \cite{Bruno:2023}), with spatial and temporal spacings given by $24/a$ and $12/a$ sites respectively. 

\begin{figure}[h]
    \centering
    \includegraphics[width=0.8\textwidth,keepaspectratio]{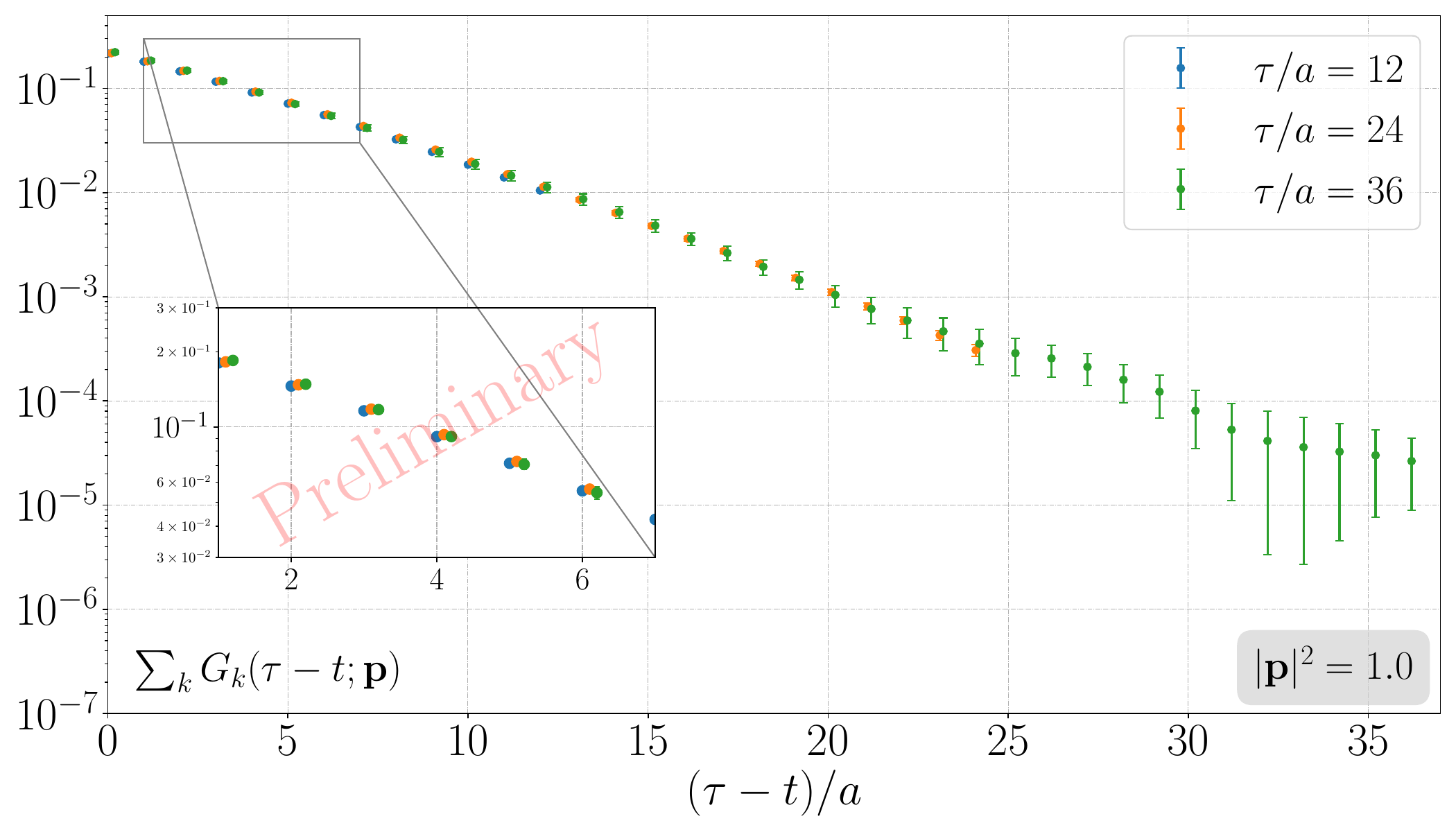}
    \caption{Two point function $G_k(\tau-t; \vb{p})$ (summed over $k$) for $\abs{\vb{p}}^2=1$ (in units of $(2 \pi/L)^2$) as a function of $\tau-t$ for different values of $\tau$. The extraction is performed according to Eq.~\eqref{eq:3pt_lim_t}, where the factor $Z_{\pi,2}^{1/2}(\vb{p})$ and the single pion energy $E_\pi(\vb{p})$ are fitted from a two-point pseudoscalar correlator. The results are blinded and the error bars are purely statistical. The points on the $x$-axis are slightly shifted for better visualization.}
    \label{fig:G2_k}
\end{figure}

The measurements of the correlators are blinded to guarantee an unbiased analysis. The different observables $C_k(t,\tau; \vb{p})$ corresponding to different momenta $\vb{p}$ (with $\tau$ fixed) have been averaged to properly construct a (temporally displaced) two-pion operator 
where only momenta with $\vb{p}_k \neq 0$ are taken into account; see Ref.~\cite{Dudek:2012gj} for an illustrative discussion on the construction of multi-hadron operators.
After computing the correlators $C_k(t,\tau; \vb{p})$, we extract the corresponding two-point function $G_k(\tau-t; \vb{p})$ from the large $t$ limit of $C_k$ while enforcing $t < \tau$, in order to select the time ordering appropriate for the time-like regime. The dependence on the low-lying pion state in Eq.~\eqref{eq:3pt_lim_t} is removed by multiplying the three-point function with the appropriate growing exponential. This situation is illustrated in the right panel of Fig.~\ref{fig:3pt_diagrams},
where the focus is on the temporal dependence of the correlator involving the operators $V_k$ and $\mathcal{O}_{\pi,1}$, namely the two-point function $G_k$. The corresponding preliminary results are shown in Fig.~\ref{fig:G2_k}, where the correlator $\sum_k G_k(\tau-t; \vb{p})$ is plotted\footnote{According to Eq.~\eqref{eq:master_formula}, the spectral density, which contains the same matrix elements of $G_k(\tau-t; \vb{p})$, is proportional to $\vb{p}_k$, therefore a Lorentz invariant combination is given by summing over the polarization index $k$ and multiplying by $\vb{p}_k$. Here, only the sum over $k$ is shown.} as a function of $\tau-t$ for the three different values of $\tau$ considered in this study. The plot highlights the expected $\tau$-independence of $G_k$ in the large $t$ limit, within the attained statistical precision.

\begin{figure}[h]
    \centering
    \includegraphics[width=0.8\textwidth,keepaspectratio]{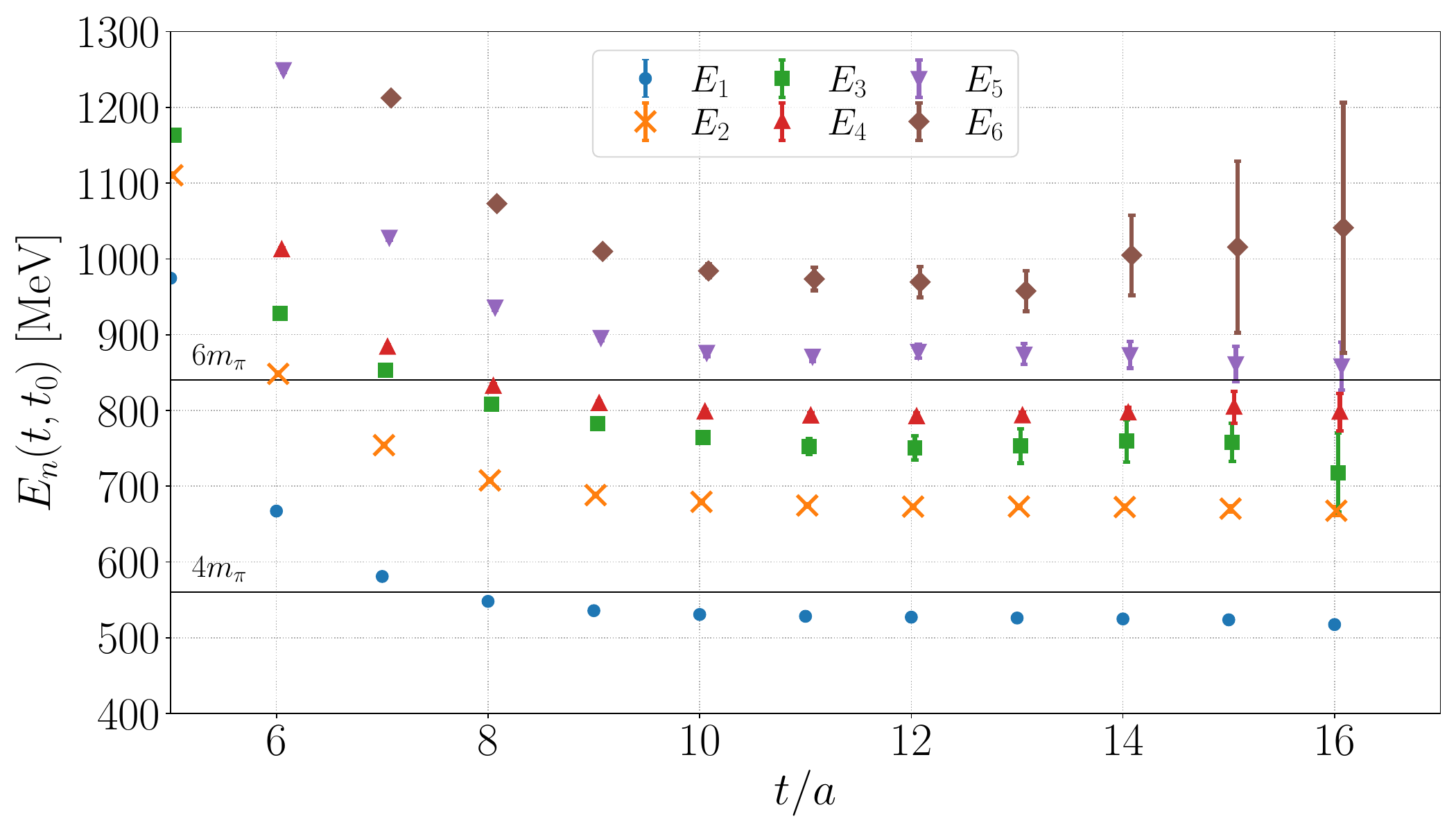}
    \caption{GEVP spectrum from the 7 operators described in the main text. The calculation of the energies has been carried out by fixing $t-t_0 = 4/a$ and by projecting the correlation matrix at each time slice onto the six eigenvectors of the one at $t_0/a = 4$, as described in Ref.~\cite{Bruno:2025mig}. The error bars are purely statistical. Points are horizontally displaced for better readability.}
    \label{fig:GEVP}
\end{figure}

As anticipated earlier, the next step is to extract the finite-volume spectral density $\rho_{k,\vb{p}}$ from $G_k$. Before doing that, the knowledge of the spectrum and the matrix elements appearing in the spectral decomposition in Eq.~\eqref{eq:spec_dec_G2} is required. To this end, based on a previous study carried out in the context of the muon $g-2$ \cite{RBC:2024fic}, a GEVP analysis on the correlator $\expval{V_k(t) V_k(0)}$ has been used to extract the energy spectrum $E_n(\vb{0})$ and the matrix elements $\mel{0}{V_k(0)}{n,\vb{0}}$. The corresponding correlator matrix is built up from a basis of $7$ interpolating operators: the local vector current $V_k$, its smeared version (in two different ways) and four smeared and polarized two-pion operators with vanishing total momentum but different relative momenta. Here we reproduced the analysis already published in \cite{Bruno:2019nzm}. The result for the spectrum is shown in Fig.~\ref{fig:GEVP}, where a clear and robust determination of the first six energy levels is observed, in the range between $0.5$ and $1$ GeV.

\begin{figure}[h]
    \centering
    \includegraphics[width=0.8\textwidth,keepaspectratio]{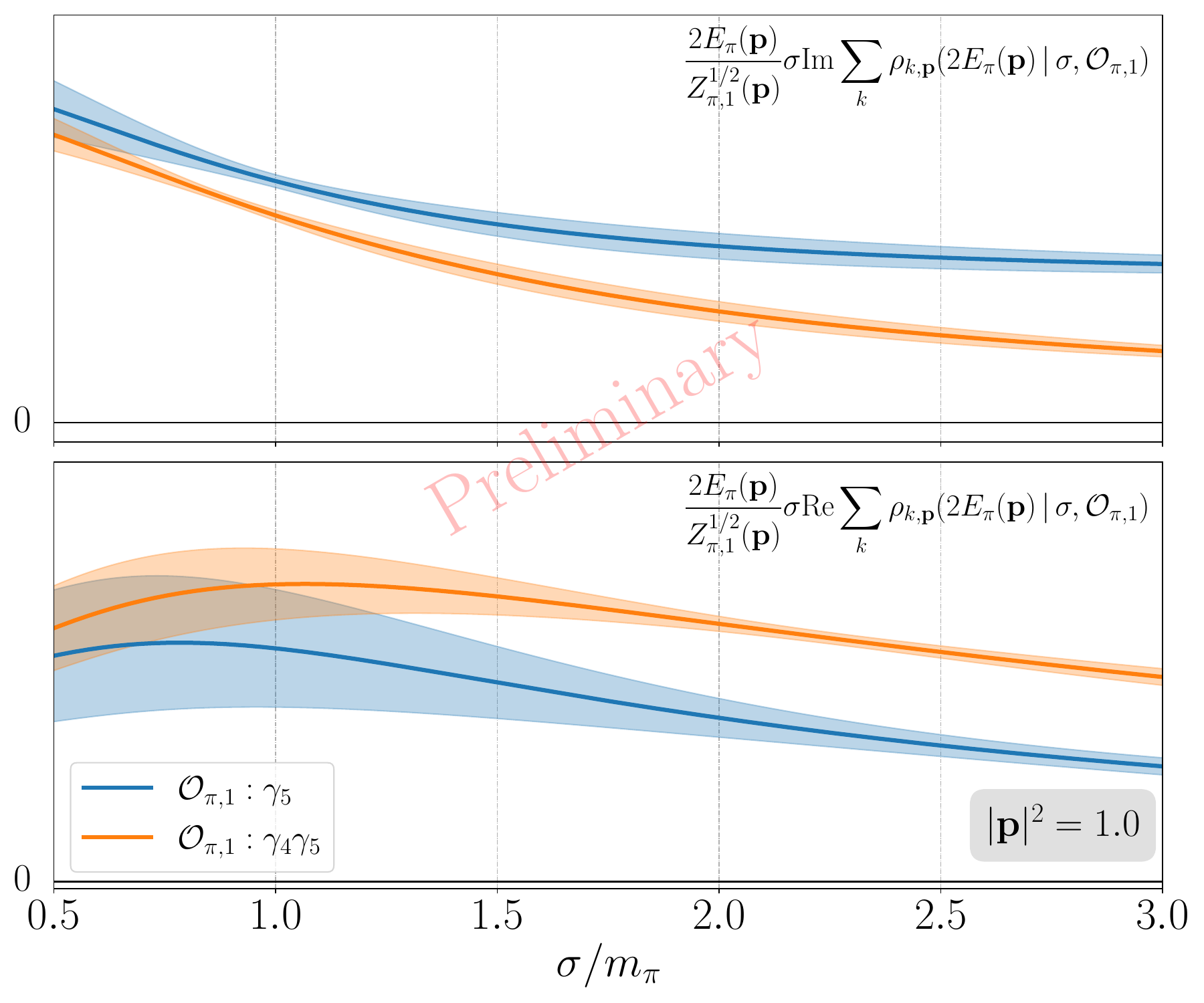}
    \caption{Reconstruction of the finite-volume spectral density $\rho_{k,\vb{p}}$ according to Eq.~\eqref{eq:rho_def} from the matrix elements $\mel{0}{V_k(0)}{n,\vb{0}}$ and energies $E_n(\vb{0})$ determined in the GEVP of Fig.~\ref{fig:GEVP}. The remaining $\mel{n, \vb{0}}{\mathcal{O}_{\pi,1}(0)}{\pi, \vb{p}}$ have been calculated by means of a constrained fit on $G_k$. The two different colors correspond to the two different choices of $\Gamma_5$ in the operator $\mathcal{O}_{\pi,1}$, see the legend in the bottom left corner. The error bars are purely statistical. \textit{Top}: reconstruction of the imaginary part. \textit{Bottom}: reconstruction of the real part.}
    \label{fig:reconstruction_rho}
\end{figure}

Once determined, we plugged $\mel{0}{V_k(0)}{n,\vb{0}}$ and $E_n(\vb{0})$ for $n=1,\dots,6$ into Eq.~\eqref{eq:spec_dec_G2} to extract the remaining matrix elements $\mel{n,\vb{0}}{\mathcal{O}_{\pi,1}(0)}{\pi, \vb{p}}$ by means of a constrained\footnote{As manifest from Eq.~\eqref{eq:rho_def}, only the product $\mel{0}{V_k(0)}{n,\vb{0}} \! \mel{n,\vb{0}}{\mathcal{O}_{\pi,1}(0)}{\pi, \vb{p}}$ for any $n$ is necessary in order to compute the associated spectral density. To this aim, our numerical investigation revealed that fixing the spectrum in the constrained fit of $G_k$ is extremely beneficial to obtain statistically acceptable fits.} fit of the correlator. Finally, in Fig.~\ref{fig:reconstruction_rho} the imaginary and real parts (top and bottom panels repectively) of the reconstructed spectral density $\rho_{k,\vb{p}}$, calculated according to Eq.~\eqref{eq:rho_def}, are plotted as a function of the smearing parameter $\sigma$. In both cases, we checked that a saturation (w.r.t. the states included in definition of the smeared density) is achieved with the first three states of the GEVP, which we use as our main result.
The plot highlights the behavior of $\rho_{k,\vb{p}}$ as a function of $\sigma$ for 2 different choices of $\Gamma_5$ in the operator $\mathcal{O}_{\pi,1}$, namely $\Gamma_5 = \gamma_5$ (blue curve) and $\Gamma_5 = \gamma_4 \gamma_5$ (orange curve). As already pointed out at the end of the previous Section, as $\sigma \to 0$ the LSZ residue is expected to be unique, therefore the two different choices of $\Gamma_5$ should yield the same limit. This is consistent with the results shown in the plot, where the two determinations become compatible for $\sigma / m_\pi \lesssim  1.0$, while in the opposite limit, the fact that they do not agree is not surprising due to the remaining "off-shellness".
However, according to Eq.~\eqref{eq:master_formula}, the limit $\sigma \to 0$ must be taken after the infinite-volume limit, which is not yet performed in this preliminary analysis. The main issue here is the delicate window problem: on one hand for $\sigma \gg 0$ the finite-volume effects are expected to be suppressed\footnote{In general, we expect finite-volume errors to be exponentially suppressed for values of $\sigma$ of the order of the mass of the pion. 
}, but the LSZ residue is not yet unique; on the other hand, for $\sigma \ll 0$ the LSZ prescription comes into play, but the finite-volume effects are no more under control.
Therefore one challenge of this study is the identification of a region where one can safely perform the extrapolation of $\sigma \to 0$ with sufficiently controlled finite-volume errors, a systematic study currently undergoing.

\section{Conclusions}\label{sec:conclusions}

In this work, we presented first exploratory tests of an inverse-problem strategy aimed at accessing QCD scattering amplitudes through a framework which exploits the LSZ reduction formula.
The preliminary calculation was performed on an ensemble at physical pion mass with Domain-Wall fermions, where the presence of the challenging window problem $1/L \ll \sigma$ and $\sigma \ll m_\pi$ makes the numerical extraction particularly delicate. 
Despite these obstacles, the observed consistency among the transition amplitudes obtained from different operators in the $\sigma \to 0$ limit provides encouraging evidence for the robustness of the method. 
Remarkably, the use of the GEVP has been pivotal for isolating the dominant states in the spectral density.

Several directions are currently being pursued to consolidate and extend these results. One objective is the direct spectral reconstruction of the two-point correlator $G_k$ using inverse-problem techniques, as described in Refs.~\cite{Hansen:2019idp,Bruno:2023bue}. In parallel, a detailed comparison using the L\"uscher finite-volume approach will serve as an important cross-check of the framework at energies (below the inelastic threshold) where both methods can be applied safely. Ongoing efforts also include increasing statistics and extending the kinematic window by exploring larger $|\vb{p}|^2$ values, together with repeating the calculation on larger lattices, while the study of discretization effects is postponed to future work.\\

{\small

\textbf{Acknowledgements.}
We thank our colleagues of the RBC and UKQCD collaborations for many valuable discussions and joint efforts over the years. We would like to thank M. T. Hansen for several stimulating discussions on this topic and for ongoing collaborations on related projects. Additionally, we thank the lattice group of the University of Milano-Bicocca for stimulating discussions. We thank CINECA for the computer time allocated via the CINECA-INFN agreements.
This work is (partially) supported by ICSC – Centro Nazionale di Ricerca in High Performance Computing, Big Data and Quantum Computing, funded by European Union – NextGenerationEU.

}

\end{document}